\documentclass[12pt]{revtex4-1}

\usepackage[utf8x]{inputenc}
\usepackage{amsbsy}
\usepackage{graphicx}
\usepackage{color}
\usepackage{dcolumn}
\usepackage{amsmath}
\usepackage{amssymb}
\usepackage{amsfonts}
\usepackage{subfigure}
\usepackage[dvipsnames]{xcolor}
\usepackage{physics}

\DeclareUnicodeCharacter{8239}{}

\newcommand{\abi}{\emph{ab initio}}
\newcommand{\bR}{\mathbf{R}}
\newcommand{\bRc}{\boldsymbol{\mathcal{R}}}

\begin{document}

\title{Superconducting hydrides on a quantum landscape}

\author{Ion Errea$^{1,2,3}$}

\affiliation{$^1$Fisika Aplikatua Saila, Gipuzkoako Ingeniaritza Eskola, University of the Basque Country (UPV/EHU),
Europa Plaza 1, 20018 Donostia/San Sebastián, Spain}
\affiliation{$^2$Centro de Física de Materiales (CSIC-UPV/EHU), Manuel de Lardizabal pasealekua 5, 20018
Donostia/San Sebastián, Spain}
\affiliation{$^3$Donostia International Physics Center (DIPC), Manuel de Lardizabal pasealekua 4, 20018 Donostia/San Sebastián, Spain}
\email{ion.errea@ehu.eus}

\begin{abstract}
Reaching superconductivity at ambient conditions is one of the biggest scientific dreams. The discoveries in the last few years at high pressures place hydrogen-based compounds as the best candidates for making it true. As the recent history shows, first-principles calculations are expected to continue guiding the experimental quest in the right track in the coming years. Considering that ionic quantum fluctuations largely affect the structural and vibrational properties of superconducting hydrides, in many cases making them stable at much lower pressures than expected, it will be crucial to include such effects on the future \abi{} predictions. The prospects for low-pressure high critical-temperature compounds are wide open, even at ambient pressure.  
\end{abstract}

\maketitle

%
%
%
%
%

\section*{A tale of success}

The discovery of superconductivity at record temperatures in pressurized hydrogen-based compounds is one of the main triumphs of theoretical and experimental physics in the last decades. The era of superconducting hydrides sparked in 2015 after the discovery of a critical temperature ($T_c$) of 203 K at 155 GPa in H$_3$S \cite{Drozdov2015}. Since then superconductivity has been reported above 200 K at megabar pressures in lanthanum \cite{PhysRevLett.122.027001,Drozdov2019}, yttrium \cite{PhysRevLett.126.117003,https://doi.org/10.1002/adma.202006832,Kong2021}, and calcium \cite{ma2021hightc} hydrogen binary compounds, as well as in C-S-H \cite{Snider2020} and La-Y-H \cite{SEMENOK202118} ternary systems. The list of superconducting hydrides continues growing, with experimental discoveries at lower but still remarkable temperatures in Th \cite{SEMENOK202036}, Pr \cite{doi:10.1126/sciadv.aax6849}, Ce \cite{PhysRevLett.127.117001}, Ba \cite{Chen2021}, Sc, and Lu \cite{doi:10.1021/acs.inorgchem.1c01960} hydrides. All these discoveries owe a great deal to the stunning developments in diamond anvil cell (DAC) techniques, which have made possible the synthesis of new hydrogen-rich compounds at high pressures and, at the same time, the study their structural and superconducting properties mainly by x-ray diffraction and resistivity measurements \cite{Flores-Livas2020,doi:10.1146/annurev-conmatphys-031218-013413}. 

Theoretical \abi{} methods also share a big part of this success story. Many of the experimental discoveries were anticipated by first-principles calculations \cite{Duan2014,PhysRevLett.119.107001,Liu6990,Wang24042012}, which predicted the thermodynamical stability at high pressures of these high-$T_c$ compounds. These calculations apply modern {\it ab initio} crystal structure prediction techniques based on density-functional theory (DFT) \cite{Flores-Livas2020,Ma_NatRevMaterials_2017,Review_Oganov-Pickard_2019}. The quest for superconducting hydrides was launched in 2004 when Ashcroft proposed that hydrogen dominant compounds may metallize and superconduct at much lower pressures than hydrogen \cite{PhysRevLett.92.187002}. Even if the first predictions right after did not yield any extraordinary $T_c$ \cite{PhysRevLett.97.045504,gao:107002,PhysRevB.76.144114,PhysRevLett.100.045504}, the persistence of some groups finally produced predictions with critical temperatures above 200 K \cite{Wang24042012}. Now hundreds of superconducting hydrogen-rich compounds have been predicted to be thermodynamically stable at high pressures \cite{Flores-Livas2020,Bi2019}, with critical temperatures ranging from few kelvins to values even above 450 K \cite{PhysRevLett.123.097001}.

With the understanding that the key to enhance $T_c$ in hydrogen-rich compounds is the creation of electronic bonding networks and a large contribution of hydrogen to the density of states at the Fermi level \cite{belli2021strong},  \abi{} calculations are expected to continue predicting new high-temperature systems, specially among ternary compounds, which still remain largely unexplored. Experimental groups will be guided, hopefully in the right direction, by these calculations and new discoveries are awaited in the coming years.

\begin{figure}[t]
\includegraphics[width=1.0\textwidth]{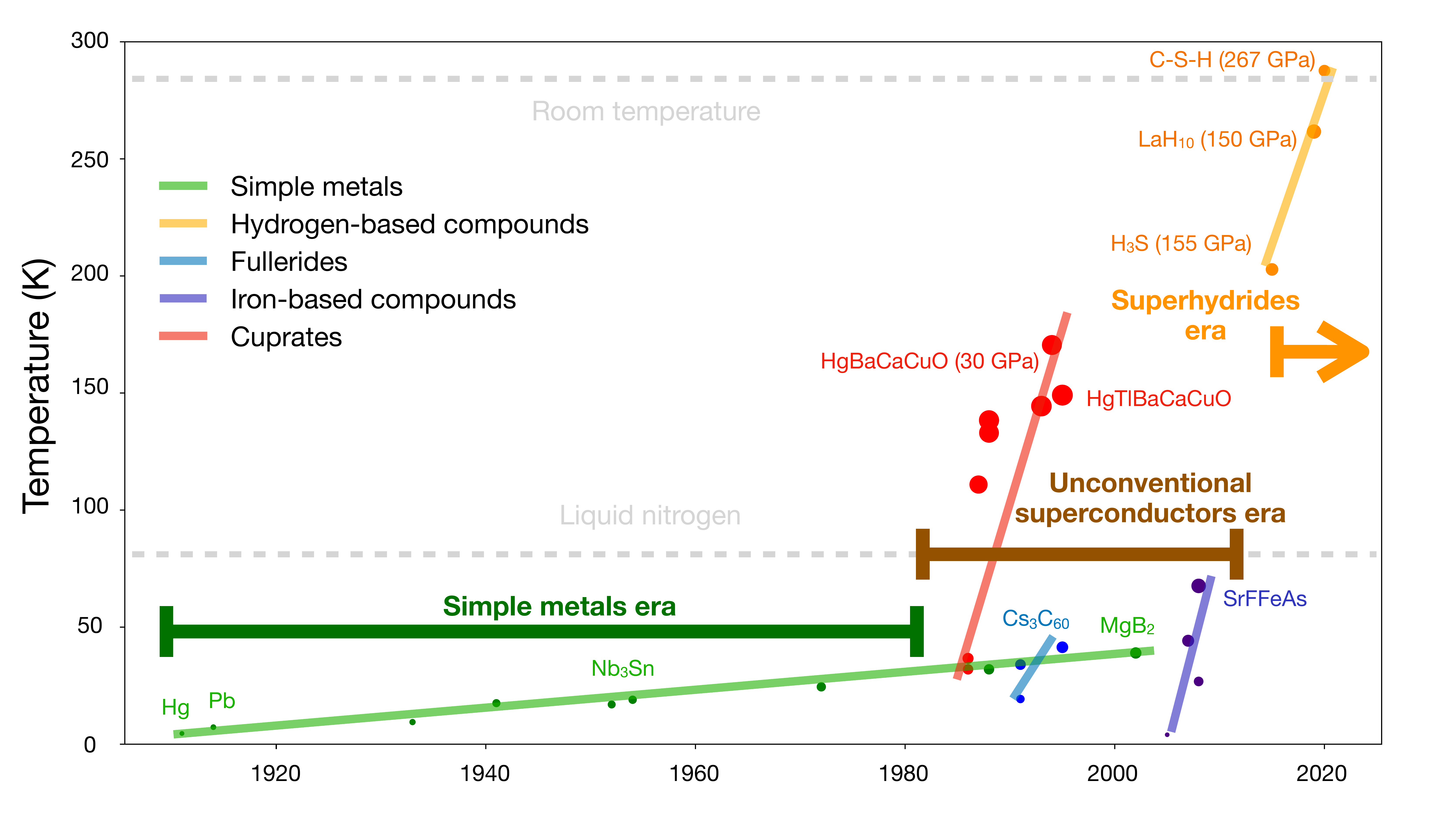}
\caption{Historical evolution of the highest superconducting critical temperatures found experimentally. Each dot corresponds to an experimentally discovered superconductor. Few of them are labeled. Different families are distinguished: simple metals, cuprates, fullerides, iron-based compounds, and hydrogen-based compounds. The size of each dot is proportional to the figure of merit as defined in Ref. \cite{doi:10.1146/annurev-conmatphys-031218-013413}, which tries to weight the extraordinary $T_c$'s of hydrogen-based compounds by the pressure at which superconductivity occurs. The eras dominated by discoveries in the simple metal, unconventional, and superhydride families are marked.}
\label{fig:tc_evol} 
\end{figure}

\begin{figure}[t]
\includegraphics[width=1.0\textwidth]{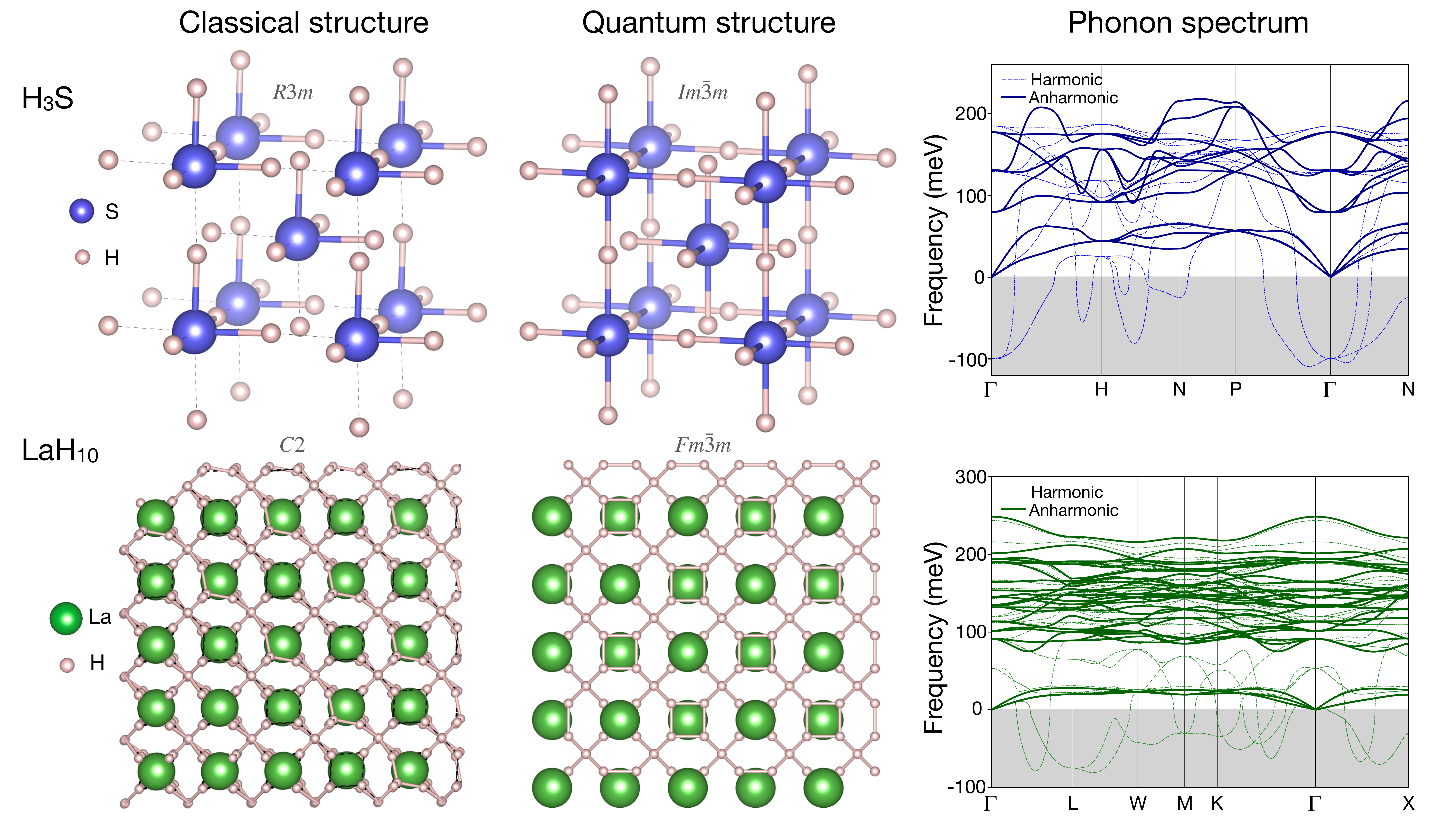}
\caption{Comparison of the crystal structures obtained in the classical approximation, neglecting the energy of ionic fluctuations, and in the quantum case, when the zero point energy is considered, for the high-$T_c$ superconducting H$_3$S and LaH$_{10}$ compounds. Quantum effects symmetrize bonds in both cases at the pressures where the highest critical temperatures are observed experimentally. The space group of each structure is given. The harmonic and anharmonic phonon spectra of $Im\bar{3}m$ H$_3$S at 135 GPa and $Fm\bar{3}m$ LaH$_{10}$ at 163 GPa are also shown. The latter data is taken from Refs. \cite{Errea81} and \cite{Errea66}.}
\label{fig:structure_phonons} 
\end{figure}

\section*{The ultimate goal}

As illustrated in Fig. \ref{fig:tc_evol}, hydrogen-rich compounds have beaten all the $T_c$ records set by the non-conventional cuprate superconductors in the early nineties. The possibility of room temperature superconductivity is real, but the problem is that this might only be true at high pressures. A good measure of the validity of a superconductor is the figure of merit $S$ introduced in Ref. \cite{doi:10.1146/annurev-conmatphys-031218-013413}:
\begin{equation}
    S = \frac{T_c}{\sqrt{T^2_{c,\mathrm{MgB}_2} + P^2}}.
    \label{eq:figure_merit}
\end{equation}
The value of $S$ of a given compound with critical temperature $T_c$ (in kelvin) is weighted by the pressure $P$ (in GPa) at which it superconducts, in comparison with the $T_c$ of MgB$_2$ ($T_{c,\mathrm{MgB}_2}=$ 39 K). Despite their large critical temperatures, the figure of merit of hydrogen-based compounds is considerably smaller than the one of the cuprates (see Fig. \ref{fig:tc_evol}), illustrating their lack of practical applications. Hydrogen-based superconductors will be useful only if they can keep their extraordinary superconducting features at ambient pressure. This poses the ultimate goal in the field and the question to be answered by the community in the coming years: is there any high-$T_c$ compound thermodynamically stable or, at least, metastable at room pressure?

There are some recent interesting theoretical predictions suggesting that there might be ternary compounds with $T_c$ values around 100 K metastable at pressures approaching ambient. LaBH$_8$ was predicted to be metastable above 40 GPa with a critical temperature of around 120 K \cite{PhysRevB.104.L020511,PhysRevB.104.134501}, while BaSiH$_8$ and SrSiH$_8$ have just been identified as potentially metastable as low as at 3 GPa and 27 GPa, respectively, with also high critical temperatures \cite{lucrezi2021insilico}. The hope that metastable high-$T_c$ phases exist at ambient pressure is thus realistic, provided that the right combination of chemical elements is found. 

\section*{A quantum energy landscape}

The dream of ambient pressure high-$T_c$ superconducting hydrides is further supported by the recent \abi{} calculations that show the crucial role of quantum ionic fluctuations in the stability of H$_3$S \cite{Errea81} and LaH$_{10}$ \cite{Errea66}, both with $T_c$ values above 200 K \cite{Drozdov2015,PhysRevLett.122.027001,Drozdov2019}. The quantum nature of the ions, which implies that due to Heisenberg's uncertainty principle ions cannot can sit steadily in a fixed lattice site and must fluctuate around it even at 0 kelvin, stabilizes the crystal structure of the phase with the highest $T_c$'s for both compounds. In fact, the high-symmetry $Fm\bar{3}m$ phase that yields the record superconductivity in LaH$_{10}$ at 250 K \cite{Drozdov2019} is stable thanks to quantum effects at least 100 GPa below the pressure at which it becomes unstable within calculations that treat the ions classically, as if they were balls that stay fixed at well-defined positions \cite{Errea66}. If quantum effects can stabilize high-$T_c$ compounds at low pressures, they become a crucial ally in the quest for ambient pressure superconducting hydrides with remarkable critical temperatures.

The difference between classical and quantum calculations is that in the former the stable and metastable crystal structures are determined  from the global and local minima of the Born-Oppenheimer energy surface $V(\bR)$, while in the latter from the minima of the quantum energy 
\begin{equation}
    E(\bRc)=\expval{T+V(\bR)}{\Psi(\bRc)}.
    \label{eq:energy}
\end{equation}$
\bR$ represents the positions of all ions in the crystal, $T$ is the kinetic energy operator of the ions, and $\ket{\Psi(\bRc)}$ the ionic wave function. We assume that the ionic wave function can be parametrized by \emph{centroid} positions $\bRc$ that determine the most probable ionic sites, i.e., the expectation value of the position operator, $\bRc = \expval{\bR}{\Psi(\bRc)}$, which would correspond to the lattice sites observed experimentally.

Fig. \ref{fig:structure_phonons} illustrates the different crystal structures obtained classically and considering quantum effects both in H$_3$S and LaH$_{10}$. The classical ground state in H$_3$S below  175 GPa is a structure with $R3m$ symmetry containing asymmetric S$-$H$\cdots$H bonds. Quantum effects symmetrize the bond making H atoms sit exactly halfway between two sulfur atoms, forming a structure with $Im\bar{3}m$ symmetry \cite{Errea81}, and keep it stable down to approximately 100 GPa \cite{PhysRevB.97.214101}. Ionic quantum effects also symmetrize the structure obtained classically for LaH$_{10}$, with space group $C2$, to a high-symmetry $Fm\bar{3}m$ phase from around 230 GPa down to at least 126 GPa. While the classical $C2$ structure consists of an irregular low-symmetry network of hydrogen atoms, the quantum $Fm\bar{3}m$ shows a fully symmetric clathrate structure. Remarkably, at the pressures at which these compounds reach the highest $T_c$'s the symmetric structures are identified by x-ray diffraction, not the distorted classical ones \cite{Drozdov2019,Einaga2016}. 

The importance of ionic quantum fluctuations can be ascribed trivially to the presence of hydrogen in the system, the lightest atom in the periodic table and, consequently, the one which is more prone to quantum fluctuations. Several other hydrogen-rich compounds \cite{https://doi.org/10.1002/adma.202006832,PhysRevLett.111.177002,PhysRevB.89.064302,PhysRevB.103.134305,doi:10.1063/5.0063968} and solid hydrogen itself \cite{0953-8984-28-49-494001,Monacelli2021Black} are also largely affected by such effects. There is, however, another less apparent but very important point that makes ionic quantum effects crucial in high-$T_c$ superconducting hydrides: their huge electron-phonon coupling induces large phonon instabilities in the harmonic approximation. Indeed, Allen and Cohen argued long time ago how a large electron-phonon interaction can soften phonon frequencies and eventually drive a system dynamically unstable \cite{PhysRevLett.29.1593}. This argument is only valid for harmonic phonons derived from the $\partial^2 V(\bR)/\partial \bR^2$ Hessian of the Born-Oppenheimer potential. In fact, as illustrated in Fig. \ref{fig:structure_phonons}, the harmonic phonon frequencies of H$_3$S and LaH$_{10}$ in the high-symmetry phases have large phonon instabilities, identified as negative eigenvalues of the Hessian, that underline that these structures are not local minima of the potential. However, in the quantum world ions live in, the stability of a structure has to be determined instead by the eigenvalues of the $\partial^2 E(\bRc)/\partial \bRc^2$ Hessian of the quantum energy. Fig. \ref{fig:structure_phonons} shows that these are indeed positive, so that the high symmetry phases are local (and also global) minima of the quantum energy. As in the calculation of $E(\bRc)$ all terms of the $V(\bR)$ potential contribute (see Eq. \eqref{eq:energy}), the latter is essentially an anharmonic result. Since high-order anharmonic terms of the Born-Oppenheimer potential will become important whenever the second-order harmonic terms are small, such as when harmonic phonons are softened, hydrides with large electron-phonon coupling are expected to be largely affected by anharmonicity.     

\section*{A complex calculation}

Calculating structural  and vibrational properties considering ionic quantum effects and the consequent anharmonicity is not a simple task for \abi{} methods, mainly due to the non-perturbative treatment that anharmonicity requires. In fact, in situations like those sketched in the phonon spectra of Fig. \ref{fig:structure_phonons} for the high-symmetry phases of H$_3$S and LaH$_{10}$ no perturbative treatment of anharmonicity is possible: the harmonic approximation has no ground state upon which a perturbative expansion can be built due to the presence of imaginary phonons. 

Despite being intrinsically non-perturbative, since the ion's dynamics are based on classical Newtonian mechanics, widely spread \abi{} molecular dynamic (AIMD) methods \cite{PhysRevLett.55.2471} fail capturing these ionic quantum effects. Phonon frequencies with the largest contribution to the electron-phonon coupling have energies of around 100 meV, which means that the ionic displacements associated to them remain essentially quantum mechanical even at temperatures as high as 1000 K. A valid alternative is offered by path integral molecular dynamics (PIMD) simulations \cite{RevModPhys.67.279}, which incorporate the quantum mechanical motion of ions at the prize of increasing the computational cost in approximately one order of magnitude. A systematic use of PIMD methods is also hindered by the difficulties of extracting phonon frequencies and identifying the equilibrium crystal structures without performing very long, and thus expensive, simulations.

A computationally less expensive alternative, which can determine the structures at the minima of the quantum energy and yield phonon spectra considering anharmonicity non-perturbatively, is offered by the stochastic self-consistent harmonic approximation (SSCHA) \cite{PhysRevB.89.064302,PhysRevB.96.014111,PhysRevB.98.024106,Monacelli_2021}. The idea of this method is to parametrize the ionic wave functions with Gaussians centered at centroid positions $\bRc$, with a width determined by some effective force constants $\boldsymbol{\Phi}$. By calculating the quantum energy in Eq. \eqref{eq:energy} and minimizing it with respect to $\bRc$ and $\boldsymbol{\Phi}$, the stable and metastable phases in the quantum energy landscape can be determined, while the phonon frequencies can be determined from the Hessian of the energy. This method has led correctly to the experimentally observed crystal structures in high-$T_c$ H$_3$S and LaH$_{10}$ \cite{Errea81,Errea66}. Including the SSCHA phonons in the calculations of the critical temperature within Migdal-Eliashberg theory, $T_c$ values in agreement with experiments have been reached as well \cite{Errea81,Errea66,PhysRevLett.114.157004}. This method has also helped to comprehend the different experimental interpretations of the metallization of high-pressure hydrogen \cite{Monacelli2021Black,Dias715,Eremets2019,Loubeyre2020}. 

\section*{Can we make crystal structure predictions on the quantum energy landscape?}

The challenge for the next years is to perform crystal structure predictions on the quantum energy landscape. All crystal structures predictions performed so far work instead on the classical Born-Oppenherimer $V(\bR)$ energy surface. It is true that these classical calculations have been able to predict  high-$T_c$ hydrides \cite{Duan2014,PhysRevLett.119.107001,Liu6990,Wang24042012}, but generally the high-symmetry and high-$T_c$ structures are found at much higher pressures than observed later experimentally due to the instabilities appearing at low pressures. If we were able to perform crystal structure predictions directly on the quantum energy landscape, we would be able possibly to predict high-$T_c$ compounds at low pressures that are currently missed by approaches not considering ionic quantum fluctuations.

The problem is that this approach requires in principle a PIMD or SSCHA calculation for each of the hundreds, or even thousands, of compounds generated in the crystal structure prediction process. This is a daunting task, not feasible with current computational resources. Discarding such an unrealistic approach, it seems more appropriate to relax in the quantum energy landscape just the most competitive phases found classically. This approach will be valid if the structures which are local minima of the classical $V(\bR)$ potential are not far from those that are minima of $E(\bRc)$, in the sense that starting, for instance, a SSCHA relaxation from the classical structures the quantum structures are obtained, like in the LaH$_{10}$ case \cite{Errea66}. 

The computational difficulties arise if the sampling of the Born-Oppenheimer potential needed to evaluate Eq. \eqref{eq:energy} is performed within DFT. As mentioned above, DFT-based crystal strucuture predictions with semi-local functionals \cite{PhysRevLett.77.3865} have turned out to be very successful. However, whether a DFT sampling is actually necessary is not clear. Could machine-learned potentials \cite{Deringer2021} alternatively be used for the quantum calculations? This would definitely speed up the quantum calculations and, provided a good potential is obtained to describe the interactions of the system, it may be possible to make structural predictions on the quantum energy landscape directly. Despite machine-learned potentials are rather sensitive to the local environment, which limits their transferability, it is encouraging that a machine learned potential has been able to describe the polymorphism of solid hydrogen at high pressures as well as its molecular to atomic liquid transition \cite{Cheng2020}. One can dream of obtaining a machine learned potential with DFT accuracy by fitting it to the forces obtained during the classical structure prediction, and later use it for the quantum relaxations also at other thermodynamic conditions.

\section*{New discoveries to come}

Considering that classical algorithms already predict metastable superconducting ternary compounds at low pressures and that ionic quantum effects seem to decrease the stabilization pressure of high-$T_c$ structures, it is reasonable to have faith on new exciting discoveries to come in the following years. The field is not at all exhausted, and the incorporation of ionic quantum effects into state-of-the-art crystal structure prediction methods will be crucial to guide experimentalists into new superconducting hydrides with a large figure of merit. Indeed, superconducting hydrides live on a quantum landscape.

\section*{Acknowledgements}

This work was supported by the European Research Council (ERC) under the European Unions Horizon 2020 research and innovation programme (grant agreement No. 802533).



\end{document}